\documentclass{ws-procs9x6}

\begin{document}

\title{Polarized Electrons for Linear Colliders\footnote{\uppercase{T}his work
 is supported by \uppercase{D}epartment of \uppercase{E}nergy contract \uppercase{DE}-\uppercase{AC}02-76\uppercase{SF}00515.}}

\author{J.~E. Clendenin, A. Brachmann, E.~L. Garwin, R.~E. Kirby, D.~-A. Luh,
 T. Maruyama, C.~Y. Prescott, J.~C. Sheppard and J. Turner}

\address{Stanford Linear Accelerator Center \\
2575 Sand Hill Road, \\
Menlo Park, CA 94025, USA \\ 
E-mail: clen@slac.stanford.edu}  

\author{R. Prepost}

\address{Department of Physics \\
University of Wisconsin, \\ 
Madison, WI 53706, USA\\ }

\maketitle

\abstracts{
Future electron-positron linear colliders require a highly polarized electron beam with a pulse structure that depends primarily on whether the acceleration utilizes warm or superconducting rf structures. The International Linear Collider (ILC) will use cold structures for the main linac. It is shown that a dc-biased polarized photoelectron source such as successfully used for the SLC can meet the charge requirements for the ILC micropulse with a polarization approaching 90\%.
}

\section{Charge}

The SLAC Linear Collider (SLC) established that reliable electron beams with a polarization at high energy approaching 80\% can be provided over periods of years. However, the beam pulse structures planned for future colliders present new demands. The ILC beam at the interaction point (IP) is expected to consist of a train micropulses spaced $\sim$300 ns apart. If one assumes that at the source it is prudent to be able to generate at least twice the charge required at the IP, then the charge, pulse length, and average current for each micropulse is indicated in Table~\ref{table1} and compared with both the Next Linear Collider (NLC) and SLC designs. The ILC linac will use superconducting (SC) L-band (LB) rf for the main linac, but normal conducting (NC) rf is a possibility for the initial acceleration including the injector. 

Highly polarized electrons for linac beams are generated by illuminating a p-doped GaAs (or its analogues) crystal with circularly polarized monochromatic light tuned to the band-gap edge. The absorbed photons promote electrons from filled valance band states to the conduction band (CB). If an atomically-clean crystal surface is treated with a Cs-oxide layer, then combined with the band-bending the work function can be lowered to below the vacuum level, resulting in a negative electron affinity (NEA) surface from which CB electrons reaching the surface are readily extracted by applying a negative bias.

The activated cathode is extremely sensitive to any contamination. To reduce field emission to near-zero, the SLC gun was operated with a bias of only -120 kV, which resulted in fields of $\sim$1.8 and 7 MV/m on the cathode crystal and electrode respectively\cite{all}. Under these conditions, space charge will limit the peak current that can be extracted to ~11 A, assuming a fully illuminated round crystal of diameter 2 cm. A practical micropulse with temporal and spatial shapes approximating a Gaussian will be limited to average currents that are significantly lower than the space charge limit (SCL); e.g., $\sim$7 A in the SLC case\cite{epp}.

If the beam generated by the source can be accelerated with the same micropulse spacing as at the IP, then Table~\ref{table1} shows there is no SCL problem regardless of whether the initial accelerating rf is SC or NC. However, if the initial accelerating rf is NC-SB to accommodate some possible damping ring designs, then the SCL could become a problem as illustrated in the 3rd column of values.

\begin{table}[ph]
\tbl{Collider charge requirements at the source and the space charge limit.}
{\footnotesize
\begin{tabular}{@{}crrrrr@{}}
\hline
{} &{} &{} &{} &{} &{}\\[-1.5ex]
{} &{} & NLC & ILC & ILC & SLC\\[1ex]
{} &{} & NC-SB & SC-LB & inj: NC-SB & Design\\[1ex]
{} &{} & {} & {} & linac: SC-LB & (2-cm)\\[1ex]
\hline
{} &{} &{} &{} &{} &{}\\[-1.5ex]
$n_e$ &nC &2.4 &6.4 &6.4 &20\\[1ex]
$\Delta t$ &ns &0.5 &2 &0.5 &3\\[1ex]
$I_{\mu pulse,avg}$ &A &4.8 &3.2 &12.8 &6.7\\[1ex]
$I_{\mu pulse,pk}$ &A &{} &{} &{} &(SCL) 11\\[1ex] 
\hline
\end{tabular}\label{table1} }
\vspace*{-13pt}
\end{table}

\section{Polarization}

The highest electron-beam polarization is achieved using crystals in which the natural degeneracy of the heavy- and light-hole energy bands at the valence-band maximum is removed. This is accomplished by introducing a lattice mismatch with the substrate or by using a short-period superlattice structure for the epilayer. By using a combination of both techniques, a separation of 50-80 meV is readily achieved. This is sufficient to tune a laser to promote electrons to the CB from the heavy-hole band only, giving promise of 100\% polarization. In practice polarizations $\geq$ 90\% have been reported\cite{wat}, and the recent parity-violating asymmetry experiment at SLAC using the 50-GeV polarized electron beam, E-158, measured $P_e=85\%$ in the early phase with a strained-layer crystal, and in the final phase an online value of $P_e=90\%$ using a strained superlattice crystal. In Table~\ref{table2}, a comparison of polarization results for 3 photocathodes representing these 2 structures is presented. The crystal operating temperature was either warm ($\sim$20$^{\circ} $ ) or cold ($\sim$0$^{\circ}$ ).

The polarization of accelerator beams is made at low energy with Mott polarimeters and at high energy with typically M\o ller or Compton polarimeters. The two offline Mott polarimeters developed for the SLC are still in use. These polarimeters gave results that were consistent with the SLC Compton polarimeter (accuracy of $<$ 0.5\%) within 2\% after corrections for known depolarization in the transmission to the IP.

\begin{table}[ph]
\tbl{Comparison of 3 photocathodes representing 2 structures.}
{\footnotesize
\begin{tabular}{@{}crrrrrr@{}}
\hline
{} &{} &{} &{} &{} &{} &{}\\[-1.5ex]
Cathode & Growth & $P_{e,max}$ & $\lambda_0 $ & $QE_{max}(\lambda_0)$ & Polarimeter & Ref \\[1ex]
Structure & Method & {} & (nm) & {} & {} & {} \\[1ex]
\hline
{} &{} &{} &{} &{} &{} &{}\\[-1.5ex]
1a GaAsP/GaAs &MOCVD &$\geq$ 0.90 &775 &0.004 &Mott &\cite{wat}\\[1ex]
strained SL &{} &{} &warm &{} &Nagoya &{}\\[1ex]
1b GaAsP/GaAs &MBE &0.86 &783 &0.012 &CTS Mott &\cite{mar1}\\[1ex]
strained SL &{} &{} &warm &{} &SLAC &{}\\[1ex]
{} &{} &0.90 &780 &0.008 &M\o ller E158-III &{}\\[1ex] 
{} &{} &{} &cold &{} &SLAC &{}\\[1ex] 
2 GaAsP/GaAs &MOCVD &0.82 &805 &0.001 &CTS Mott &\cite{mar2}\\[1ex] 
strained-layer &{} &{} &warm &{} &SLAC &{}\\[1ex]
{} &{} &0.85 &800 &0.004 &M\o ller E158-I &\cite{ant}\\[1ex] 
{} &{} &{} &cold &{} &SLAC &{}\\[1ex] 
\hline
\end{tabular}\label{table2} }
\vspace*{-13pt}
\end{table}

Since the electron polarization can vary significantly from one crystal to another for the same crystal structure, and can even vary by a few percent for a single crystal depending on surface conditions, it has been very difficult to compare the polarization results between different facilities. Some insight into the accuracy of the polarimeters themselves was recently gained in an experiment at JLAB in which the polarized electron beam was switched on a pulse-to-pulse basis between 3 different M\o ller polarimeters and a Compton polarimeter, all at high energy, and compared with an inline Mott polarimeter at low energy\cite{gra}. Undetected systematic effects and errors in the simulated values of the effective analyzing powers were revealed at the level of several percent. 

As shown in Table~\ref{table2}, the higher polarization using cathode 1b relative to cathode 2 is the same when measured by 2 independent polarimeters. If the E158 M\o ller polarization data---corrected for the 1\% depolarization between source and polarimeter---for a GaAsP/GaAs strained superlattice crystal is combined with the SLAC Mott measurements made with a cathode crystal cut from the same wafer, the result is a polarization at the source of (88$\pm$ 4)\%.

\end{document}